\documentclass[]{agujournal2019}
\usepackage{amsmath,amssymb}
\journalname{Geophysical Research Letters}

\begin{document}

%
%


\title{Observational constraint on the radius and oblateness of the lunar core-mantle boundary}

%
%




\authors{V. Viswanathan\affil{1,2}, N. Rambaux\affil{1}, A. Fienga\affil{1,2}, J. Laskar\affil{1}, M. Gastineau\affil{1}}


\affiliation{1}{ASD/IMCCE, Observatoire de Paris, PSL Universit\'e, Sorbonne Universit\'e, 77 avenue Denfert-Rochereau, 75014 Paris, France}
\affiliation{2}{AstroG\'eo/G\'eoazur -- CNRS/UMR7329, Observatoire de la C\^ote d'Azur, 250 rue Albert Einstein, 06560 Valbonne, France}




\correspondingauthor{Vishnu Viswanathan}{vishnu.viswanathan@obspm.fr}
\correspondingauthor{Nicolas Rambaux}{nicolas.rambaux@obspm.fr}






\begin{keypoints}
\item LLR and GRAIL gravity field data are used to perform fits of a lunar interior's dynamical model.
\item Estimates of core oblateness intersect with corresponding theoretical values of a hydrostatic core.
\item The accuracy of oblateness and radii of a presently-relaxed lunar core is improved by a factor of 3.
\end{keypoints}

%
%


\begin{abstract}
Lunar laser ranging (LLR) data and Apollo seismic data analyses, revealed independent evidence for the presence of a fluid lunar core. However, the size of the lunar fluid core remained uncertain by $\pm55$ km (encompassing two contrasting 2011 Apollo seismic data analyses).
Here we show that a new description of the lunar interior's dynamical model provides a determination of the radius and geometry of the lunar core-mantle boundary (CMB) from the LLR observations. We compare the present-day lunar core oblateness obtained from LLR analysis with the expected hydrostatic model values, over a range of previously expected CMB radii.
The findings suggest a core oblateness ($f_c=(2.2\pm0.6)\times10^{-4}$) that satisfies the assumption of hydrostatic equilibrium over a tight range of lunar CMB radii ($\mathcal{R}_{CMB}=381\pm12$ km).
Our estimates of a presently-relaxed lunar CMB translates to a core mass fraction in the range of $1.59-1.77\%$ with a present-day Free Core Nutation (FCN) within $(367\pm100)$ years.

\textbf{Plain language summary}
The study of the rotation of a body gives access to key information about its interior. Using a set of numerically-integrated equations and the knowledge of Moon's gravity from the GRAIL mission, we are able to simulate the rotation and motion of the Moon in the vicinity of Earth, Sun and other planetary bodies with high accuracy. In this study, we compare the expected relaxed shape of the Moon's core with that obtained from a best-fit adjustment of our simulation parameters to the observed LLR data. This novel approach allows us to improve the previous uncertainty in the radius and polar flattening of the Moon's core-mantle boundary (CMB), both by a factor of 3. Limits on the size of the lunar CMB provide significant constraints to important works such as the Earth-Moon formation (e.g. giant impact) hypotheses. In addition, a better constraint on the lunar CMB radii translates to an improvement on the precision tests of fundamental physics using LLR data. Furthermore, our methods can be applied to study the influence of the liquid core on the rotation of other planets, especially Mars, with the recent advent of the InSight mission.
\end{abstract}
\section{Introduction}
\subsection{State of the art}
Lunar laser ranging (LLR) consists of measuring the round-trip travel time of a laser pulse emitted from an observing station on the Earth and received back after bouncing-off of a retro-reflector array on the surface of the Moon. LLR observations to these optical devices on the near-side of the Moon (five sites, as a part of the payloads of the NASA Apollo and Russian Lunokhod missions) continue to be collected since 1969 \cite{Bender1973}. The accuracy of these range measurements gradually improved from the initial few tens of centimeters in the 1970s, to a few centimeters in the 1990s, to millimeter-level accuracies since the 2000s \cite{Murphy2013,Courde2017}.
At present, the entire LLR dataset span 48 years in time, greater than a factor of 2 times the period of lunar nodal precession of 18.6 years. The analyses and results retrieved by using such highly accurate range measurements, span multi-disciplinary science such as geodesy and geodynamics, solar system ephemerides, terrestrial and celestial reference frames, lunar physics and fundamental physics (e.g., \citeA{Dickey1994,Murphy2013}).

The lunar science derived from LLR depends on the accurate monitoring of the time-varying lunar orientation and orbital motion. A mathematical description of the orbital and rotational dynamics of the Moon is referred to as the dynamical model. This model includes the mutual interactions between the interior layers of the Moon (i.e., crust/mantle and fluid core) as well as perturbations from other planetary bodies. They also describe the lunar orientation through Euler angles and state vectors, which are fitted to the reduced LLR observations (see Appendix). The combination of LLR observations with the lunar gravity-field solutions derived from the GRAIL mission \cite{Konopliv2013,Lemoine2013} allows strong constraints to be placed on the dynamical model, enabling LLR to better resolve some correlated model parameters \cite{Williams2014b,Pavlov2016,Viswanathan2018}. The INPOP17a \cite{Viswanathan2018} version allowed us to compare and validate our lunar dynamical model, LLR reduction procedure and parameter adjustments against other analysis groups.

\subsection{Context}
The presence of a fluid core alters the angular momentum balance between the layers modeled through the Euler-Liouville equations for the total Moon (or the Moon),
\begin{equation}
 \frac{d}{dt}(\mathcal{I}\boldsymbol{\Omega} + \mathcal{I}_c\boldsymbol{\omega_c}) + \boldsymbol{\Omega} \times (\mathcal{I}\boldsymbol{\Omega} + \mathcal{I}_c\boldsymbol{\omega_c}) = \boldsymbol{\Gamma}^{external}\end{equation}
 and for the fluid core,
 \begin{equation}
 \frac{d}{dt}\mathcal{I}_c(\boldsymbol{\Omega + \omega_c}) + \boldsymbol{\Omega} \times \mathcal{I}_c(\boldsymbol{\Omega + \omega_c}) = \boldsymbol{\Gamma}_c^{friction} + \boldsymbol{\Gamma}_c^{inertial}.\end{equation}
 Here, $\mathcal{I}$ is the moment of inertia (MoI) tensor for the Moon, $\boldsymbol{\Omega}$ is the angular velocity of the Moon and $\boldsymbol{\Gamma}^{external}$ is the sum of the external torques acting on the Moon (i.e., figure-point mass interactions, figure-figure interactions and de Sitter precession). The subscript ``c'' represents equivalent parameters for the core. We define $\boldsymbol{\omega}_c$ as the angular velocity of the lunar core relative to that of the Moon. The lunar coordinate system is defined by the principal axes of the undistorted Moon, where the MoI tensor is diagonal. A set of Euler angles ($\phi$, $\theta$, $\psi$) defines the orientation of the principal axes frame to the inertial (ICRF2) frame. The MoI of the Moon varies with time due to tidal distortions from the Earth, Sun and spin distortion \cite{Viswanathan2018}. The component of permanent-tide is included within the tidal and spin-distortions \cite{Williams2001a}. The modeled dissipative torques arise from viscous friction due to differential rotation \cite{Folkner2014} at the CMB ($\boldsymbol{\Gamma}_c^{friction}$), while the inertial coupling torques \cite{Rambaux2007} ($\boldsymbol{\Gamma}_c^{inertial}$) arise from the flow of the fluid along a non-spherical CMB.

 The exchange of angular momentum between the layers forms the basis of sensitivity of LLR to the size and shape of the fluid core.

\subsection{Motivation}
LLR solutions are non-unique to a range of fluid core sizes (e.g. Fig. \ref{img_residuals}) and this non-uniqueness primarily arises from model parameter correlations in the fit (see Section \ref{sec:Method}). This study shows that the lunar core's hydrostatic nature can be used as an apriori to improve the previous constraints on the Apollo-seismic data determined radius and LLR-observed geometry of the lunar core-mantle boundary (CMB), both by a factor of 3 (see Section \ref{sec:Results}). We show that this improvement allows a better determination of some derived quantities (e.g. lunar core mass fraction and lunar free core nutation) followed by concluding remarks on the future perspectives and applicability of this method to other planets.

\section{Methodology}
\label{sec:Method}
The LLR model is compatible with a range of fluid core sizes \cite{Williams2014b}, often represented by the value of the ratio of the polar moment of inertia of the lunar core to the total Moon ($\alpha_c=\mathcal{C}_c/\mathcal{C}_T$).
The previous solution INPOP17a \cite{Viswanathan2018} fixed $\alpha_c$ to a model value ($7 \times 10^{-4}$), primarily due to its correlation (Pearson correlation coefficient of -0.8) with the core oblateness $f_c$ (where $f_c = [\mathcal{C}_c - (\mathcal{A}_c + \mathcal{B}_c)/2]/\mathcal{C}_c$ is used to describe the polar flattening of the core, through its principal components of the moment of inertia tensor $\mathcal{A}_c$, $\mathcal{B}_c$ and $\mathcal{C}_c$). While this allowed a close comparison to independent studies \cite{Folkner2014,Pavlov2016}, the previously reported \cite{Viswanathan2018} uncertainty on $f_c$ does not account for uncertainties from considering a fixed value for $\alpha_c$. A different plausible model value of $\alpha_c$ (e.g., $3\times10^{-4}$) would increase the corresponding value of $f_c$ by $\approx 2\times10^{-4}$, suggesting an uncertainty $\delta f_c \approx \frac{2\times10^{-4}}{\sqrt{2}}\approx\pm 1.4\times10^{-4}$ (see \citeA{Williams2014b}). The uncertainty of $f_c$ obtained thereof, encompasses the range of $f_c$ obtained for a fluid core radii varying from $\approx$ 320 to 440 km (see Fig.~\ref{img_fc}).

In a more direct approach, this study used the radius of the lunar core-mantle boundary ($\mathcal{R}_{CMB}$) as a model parameter, by redefining the principal components of the moment of inertia of the core ($\mathcal{A}_c$, $\mathcal{B}_c$ and $\mathcal{C}_c$). Other geophysical parameters involved in the moment of inertia redefinition include the mean core density ($\rho_c$) and the CMB shape coefficients ($d_{nm,c}$, $e_{nm,c}$ where $n, m$ are the degree and order) to represent a triaxial core, given by,
\begin{equation}\label{MoI_c}\centering
        \mathbf{\mathcal{I}_c} =    \frac{8\pi\rho_c\mathcal{R}_{CMB}^5}{15\mathcal{M}\mathcal{R}_T^2} \begin{bmatrix}
                  1+\frac{1}{2}d_{20,c}-3d_{22,c}     &-3e_{22,c}         &-\frac{3}{2}d_{21,c}    \\
                  -3e_{22,c}         &1+\frac{1}{2}d_{20,c}+3d_{22,c}     &-\frac{3}{2}e_{21,c}    \\
                  -\frac{3}{2}d_{21,c}         &-\frac{3}{2}e_{21,c}         &1-d_{20,c}
                                  \end{bmatrix}.
      \end{equation}
The CMB polar shape coefficient ($d_{20,c}$) can be represented in terms of the core oblateness ($f_c$) using the integrals of the principal moments, given by $d_{20,c}=(-2/3)f_c$ \cite{Meyer2011}.
This representation \cite{Richard2014a} (Eqn.~\ref{MoI_c}) is convenient to explore plausible values of the moment of inertia of the lunar core and place constraints on itself through a range of lunar interior models with varying core radii, densities and surface shape coefficients. A set of equations based on this representation was implemented within INPOP, considering a lunar crust, mantle and a triaxial fluid core. The triaxiality of the lunar core introduces additional components to the inertial coupling torque expansion (e.g. \citeA{Rambaux2007}) that impact the rotation of the Moon.

A reference lunar interior model is built from INPOP17a \cite{Viswanathan2018} parameters and consists of three layers (crust, mantle and fluid core) of constant density. For a given core radius, the reference (or hydrostatic CMB with non-hydrostatic lithosphere) model provides constraints on the core density and shape. The shape of the CMB for the reference model is calculated from a combination of the gravitational attraction of the crust, mantle, centrifugal acceleration, and mean tides (e.g. \citeA{Meyer2011,Dumberry2016,Wieczorek2019}). A more detailed discussion on the reference model (e.g. \citeA{Chambat2008,Weber2011,Garcia2011,Williams2014b,Antonangeli2015,Matsuyama2016}) can be found in the Supporting Information (SI). The gravity field of the Moon is constrained up to degree and order-6 from a GRAIL analysis \cite{Konopliv2013}. With the help of these constraints, an iterative least-square fit of the lunar dynamical model parameters to LLR data is performed. Each iterative fit started with initial values of geophysical parameters ($d_{20,c}$, $\mathcal{R}_{CMB}$ and $\rho_c$) from the hydrostatic model. Subsequent iterations in the fit allowed for deviations of $d_{20,c}$ (the parameter of interest in this study) from the corresponding initial hydrostatic value. The fit of $d_{20,c}$ was necessary to maintain the recent (and most accurate) LLR post-fit weighted root-mean-square (wrms) to well-below 2 cm (see Fig.~\ref{img_residuals}). The value of the CMB equatorial shape coefficient ($d_{22,c}$) was held fixed to its hydrostatic value during the iterations, due to its insufficient sensitivity in the fit. Fits to LLR data show that the impact of varying the value of $d_{22,c}$ is indistinguishable at the present level of data accuracy. However, we still take into account a non-zero value of $d_{22,c}$ to quantify its impact on the estimation of $d_{20,c}$.

The off-diagonal elements of the moment of inertia of the core (containing surface coefficients $d_{21,c}$, $e_{21,c}$ and $e_{22,c}$ in Eqn.~\ref{MoI_c}) are set to zero to align the principal moments of the lunar core with the principal axes of the undistorted Moon. \citeA{Wieczorek2019} show possible deviations from this perfect alignment when a non-hydrostatic lithospheric model is considered, giving about $6.4^{\circ}$ of tilt between the core principal polar moment with that of the Moon. We show that such a misalignment would introduce a relative error of below $1\%$ on our core oblateness estimates (see SI).

\section{Results and Discussions}
\label{sec:Results}
\begin{figure}
\includegraphics[width=1.0\textwidth]{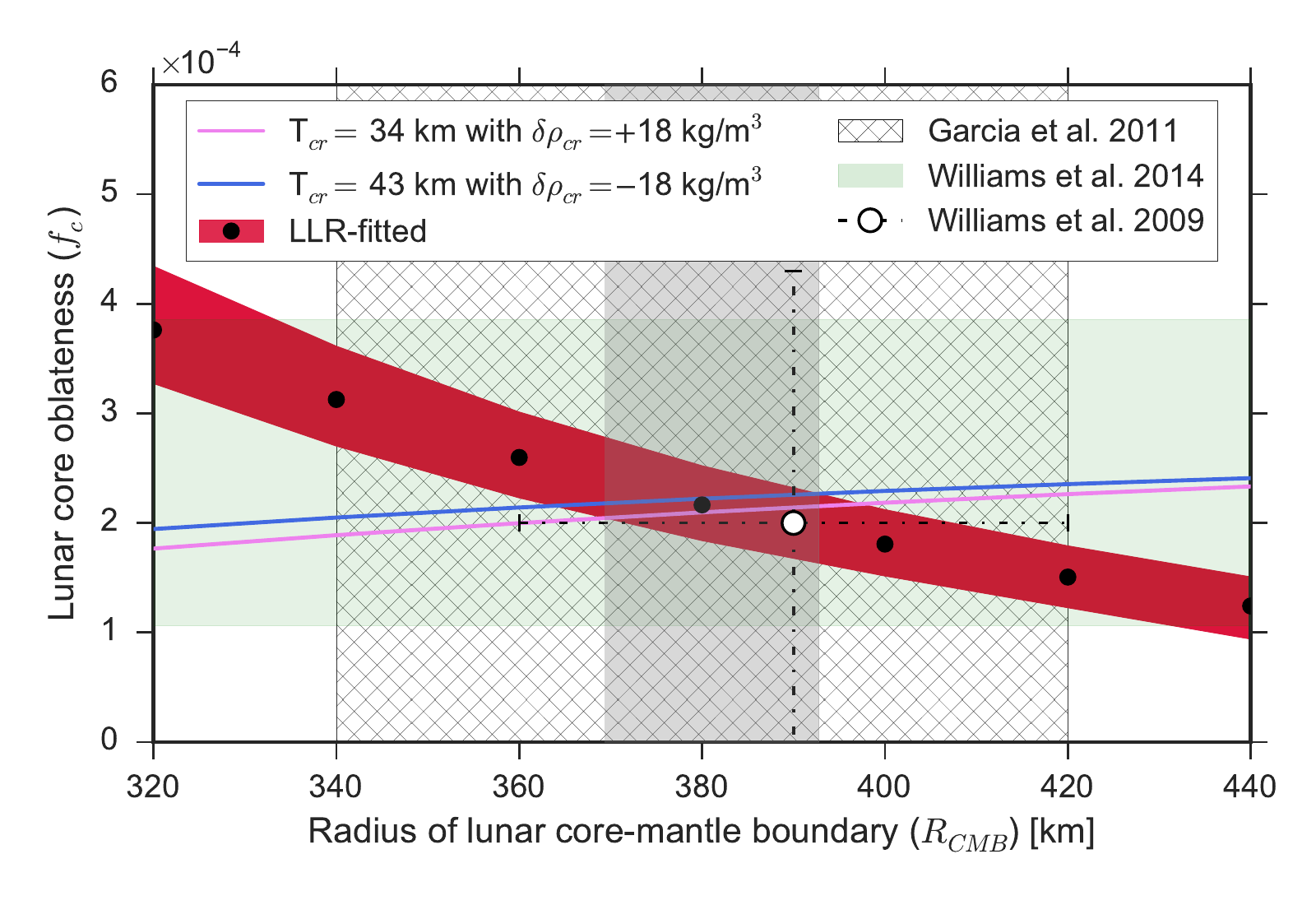}
\caption{The LLR-fitted value of the lunar core oblateness $f_c$ (in black dots with region of uncertainty in red) intersects the theoretical hydrostatic values of $f_c$ (solid lines in violet and blue corresponding to models with two different lunar crustal thicknesses ($34$ and $43$ km) with $\pm 18$ kg/m$^3$ crustal density variations, respectively) at a lunar CMB radius of $\mathcal{R}_{CMB}=381 \pm 12$ km (in gray region).
The LLR-fitted mean values here are obtained by assuming a mean value of lunar crustal thickness (T$_{cr}=(34+43)/2=38.5$ km) and density ($\rho_{cr}=2550\pm18$ kg/m$^3$) estimates \cite{Wieczorek2013} in the LLR dynamical model. A model with T$_{cr}=43$ km and $\delta \rho_{cr}=-18$ kg/m$^3$ tends to increase the LLR-fitted mean value of $f_c$ by $10.9$ to $7.7\%$, while a T$_{cr}=34$ km and $\delta \rho_{cr}=+18$ kg/m$^3$ tends to decrease the same by $10.7$ to $8.5\%$, for $\mathcal{R}_{CMB}$ varying from 320 to 440 km, respectively.
The region of uncertainty of the LLR-fitted $f_c$ (in red region) encompasses the cumulative errors from lunar core density \cite{Garcia2011}, crustal thickness and mean density variations \cite{Wieczorek2013}, degree-2 potential Love number \cite{Konopliv2013}, and other parameters listed in Table \ref{table_fc_sig} in the order of decreasing precedence. Previously reported \cite{Williams2009} $f_c$ ($2.0\pm2.3\times10^{-4}$) is in agreement but with much larger error bars (in white dot). A more recent estimate \cite{Williams2014b} ($2.42\pm1.4\times10^{-4}$) covers plausible values of $f_c$ obtained for $\mathcal{R}_{CMB} \approx 320$ to $440$~km (in green region).
The estimated value of $\mathcal{R}_{CMB} = 381 \pm 12$ km (in gray region) is obtained by the intersection of the lower and upper bounds of LLR-fitted $f_c$ with the hydrostatic models of T$_{cr}=34$ and $43$ km, respectively (see SI). The CMB radius agrees within 1-$\sigma$ of the Apollo seismic data analysis by \citeA{Garcia2011} (in hatched region) and differs by $13 \%$ with \citeA{Weber2011}. Within these limits, the value of lunar core oblateness ($f_c$) is estimated as $(2.2 \pm 0.6) \times 10^{-4}$.}
\label{img_fc}
\end{figure}

The CMB polar shape coefficient ($d_{20,c}$) is fitted to LLR data over a range of previously expected CMB radii. Fig.~\ref{img_fc} shows this fitted value expressed in terms of the core oblateness ($f_c$) to allow comparisons with previous LLR estimates. The range of LLR-fitted (observed) values of $f_c$ crosses its corresponding theoretical hydrostatic values (obtained by considering variations in crustal thickness ($34$ and $43$ km) and average density ($2550 \pm 18$ kg/m${^3}$) from \citeA{Wieczorek2013}) at a CMB radius of $381 \pm 12$ km (highlighted region in gray). Within these limits, we obtain an estimated value of lunar core oblateness $f_c = (2.2 \pm 0.6) \times 10^{-4}$.

The intersection of the observed and theoretical values of $f_c$ signifies that at the level of sensitivity of LLR datasets, the present-day lunar fluid core geometry satisfies the theoretical considerations of the case of a hydrostatic lunar fluid core within a non-hydrostatic lunar lithosphere. This is a suggested observational evidence in agreement with previous model predictions \cite{Meyer2011,LeBars2011}.
A recent study that used three-dimensional mantle convection models \cite{Zhang2017} suggests the presence of a partially molten ilmenite-bearing cumulates (IBCs) rich layer with low-viscosity surrounding the present-day lunar core.
This offers additional explanation to the previously proposed low-viscosity, seismically attenuating layer near the CMB \cite{Khan2014,Harada2016}. With the viscosity at the base of the mantle $\eta_b~\approx10^{19}$ Pa s (a conservative value of overturned IBCs with 6 wt\% ilmenite \cite{Zhang2017}), an approximate order of relaxation timescales of the CMB can be computed using the expression $\tau_r \sim \eta_b\mathcal{R}^2_{CMB}/(\rho_c\Delta\rho\mathcal{G}\delta^3)$ \cite{Nimmo2012}. Here, $\eta_b$ is the viscosity at the base of the lunar mantle ($\sim 10^{19}$ Pa s from \citeA{Zhang2017}), $\Delta\rho$ is the density contrast between the lunar core and mantle ($\sim$ 2500 kg/m$^3$), and $\delta$ is the temperature and activation energy dependent effective channel thickness ($\sim 21$ km). With these conservative choice of values, we obtain CMB relaxation timescales of up to a few tens of Myr (compared to the $\sim 4.5$ Gyr time since the formation of the Moon), supporting a present-day hydrostatic (or relaxed) core \cite{Meyer2011,LeBars2011} within a frozen-in non-hydrostatic lithosphere (e.g. \citeA{Garrick-Bethell2014}).

The region of error on the observed values of $f_c$ was obtained after considering the impact of correlated parameters (fixed or constrained from previous analyses) in the fit (see Appendix). The largest contribution to the uncertainty on $f_c$ ($\le 20\%$) arises from the uncertainty of the lunar fluid core density (between $\approx 5000$ to $7000$ kg/m$^3$) from the analysis of Apollo seismic data. The range of CMB radii shown in Fig.~\ref{img_fc} yield equally good fits of the lunar dynamical model to the LLR data used. This is evident from the variations of the weighted root-mean-square of LLR post-fit residuals (see Fig.~\ref{img_residuals}), obtained after iterative fits of models with varying core radii, at an order of magnitude below the present-day LLR observational accuracy of about $5$ mm \cite{Murphy2013,Courde2017}.

The mean value of the radius of the lunar CMB satisfying both the observed and the theoretical values of the core oblateness agree at a relative error of 0.3\% with the Apollo seismic data analysis by \citeA{Garcia2011}, and differ by 13 \% with \citeA{Weber2011}. The improvement in the uncertainty with respect to \citeA{Garcia2011} is by a factor of 3 to the hydrostatic case. Our estimates of  CMB radii are in agreement with the analysis of Lunar Prospector spacecraft's magnetometer measurements \cite{Hood1999}.

The mass of the lunar core that corresponds to the estimated range of hydrostatic CMB radii lies between $1.59$ to $1.77$\% of the total lunar mass $\mathcal{M}$ (where $\mathcal{M} \approx7.346 \times 10^{22}$ kg is derived from joint lunar and planetary fits \cite{Viswanathan2018}). Previous estimates are in close agreement and lie between $1-3\%$ \cite{Hood1999} from Lunar Prospector (LP) mission, $\le1.5\%$ \cite{Williams2014b} from GRAIL mission and $1.7-2.5\%$ \cite{Rai2014} from the analysis based on siderophile element content in the lunar mantle.
Simulations of moon-forming impact collisions \cite{Canup2001} used upper-limits on the previously estimated mass fraction ($1-3$\%) \cite{Hood1999} of the present-day lunar core as a proxy for the mass fraction of iron in the orbiting equatorial disk mass ($M_{Fe}/M_D$), expected as a consequence of a giant impact on the proto-Earth. This enables the core mass fraction to constrain a range of head-on to off-axis collisions considered by such studies \cite{Canup2001,Canup2012}.

The Free Core Nutation (FCN) is a mode related to the non-alignment of the axis of rotation of the core and the mantle. The period of the FCN of the Moon (in days) is related to the core oblateness approximately as $\mathcal{P}_{FCN}\approx27.32/f_c$ \cite{Rambaux2011}. This gives a present-day $\mathcal{P}_{FCN} \approx (367 \pm 100)$ years for the hydrostatic case, assuming a Poincar\'e flow within the lunar fluid core. The large value of $\mathcal{P}_{FCN}$ with respect to the mantle precession (18.6 years) confirms that the present-day lunar core should be decoupled with the mantle \cite{Meyer2011}.

Tests of fundamental physics using LLR data rely on the accuracies of both the measurement and the model.
Inaccurate size of the modeled lunar core introduces systematic biases in the tests of the principle of universality of free fall, estimated using parameter adjustments to LLR data (see a previous demonstration \cite{Williams2012b}). We validate the origin of such biases (of $\approx 4 \times 10^{-14}$ on the mean value of the fractional differential acceleration of the Earth and the Moon towards the Sun) by using plausible values of $\alpha_c$. We obtain similar differences in solution values as given by \citeA{Williams2012b}. Such biases are significant to LLR tests of the universality of free fall, since the current LLR detection limit is at the level of $\approx 7 \times 10^{-14}$ \cite{Viswanathan2018}.

\section{Conclusions and Perspectives}
This study compares the present-day lunar core oblateness obtained from LLR analysis with the expected hydrostatic model values, over a range of previously expected CMB radii. The findings suggest a core oblateness ($f_c=(2.2 \pm 0.6) \times 10^{-4}$) that satisfies the assumption of hydrostatic equilibrium over a tight range of lunar CMB radii ($\mathcal{R}_{CMB} = 381 \pm 12$ km). This range of CMB radii agrees within one-$\sigma$ of both seismological analysis \cite{Garcia2011} and spacecraft magnetometer analysis \cite{Hood1999}. The accuracy of $f_c$ and $\mathcal{R}_{CMB}$ is improved by a factor of 3.
Our estimates of a presently-relaxed lunar CMB translates to a core mass fraction in the range of $1.59-1.77\%$, a parameter to limit the possible scenarios of giant-impact during the formation of the Moon \cite{Canup2001,Canup2012}. The estimated core oblateness causes the present-day Free Core Nutation (FCN) of the Moon to be within $(367 \pm 100)$ years. Furthermore, an improvement in the knowledge of the lunar core radii allows a better understanding of the systematic biases in the solution values of LLR equivalence principle (EP) tests.

Future extension of the Apollo seismometer network with a better coverage \cite{Mimoun2012} would allow a better determination of $\rho_c$ and $\mathcal{R}_{CMB}$ thereby improving current LLR estimations. With advancements in the LLR measurement \cite{Courde2017,Adelberger2017} continuing to accumulate high-accuracy datasets and emerging observational techniques \cite{Dehant2017}, future LLR analysis will allow unprecedented access to the dynamical nature of the lunar interior. Moreover, the methods described here can be applied to study the influence of the liquid core on the rotation of other planets such as Mars \cite{Folkner2018}.
\appendix
\section{Lunar interior model description}
\subsection{Dynamical model}
The dynamical equations within INPOP \cite{Viswanathan2018} consider a uniform density lunar fluid core with its rotation resembling a rigid body and whose shape and size are constrained by the CMB.
The expression for the moment of inertia (MoI) of the undistorted total Moon can be expressed as:
\begin{linenomath*}
\begin{equation}
      \label{MoI_Moon}
      \mathcal{I}_T =  \dfrac{\mathcal{C}_T}{\mathcal{M}\mathcal{R}^{2}_{T}}\begin{bmatrix}
                                                         1 &0 &0 \\
                                                         0 &1 &0 \\
                                                         0 &0 &1\end{bmatrix} +
          \begin{bmatrix}
                  C_{20,T} - 2C_{22,T}      &0                                          &0    \\
                  0                                           &C_{20,T} + 2C_{22,T}     &0    \\
                  0                                           &0                                          &0
          \end{bmatrix}
      \end{equation}
\end{linenomath*}
where C$_{20,T}$ and C$_{22,T}$ are the unnormalized degree-2 Stokes coefficients for the spherical harmonic model of the undistorted Moon and $\mathcal{C}_T/\mathcal{MR}^{2}_T$ is the undistorted polar MoI of the Moon normalized by its mass $\mathcal{M}$ and radius squared $\mathcal{R}^{2}_T$. Through Eqn.~\ref{MoI_Moon}, we can directly use the undistorted value of $C_{20,T}$ and $C_{22,T}$ from GRAIL-derived gravity field models \cite{Konopliv2013}.

The moment of inertia matrix for a core ($\mathcal{I}_c$) can be represented using the Eqn.~\ref{MoI_c} where polar component $\mathcal{C}_c$ can be expressed equivalently in terms of lunar geophysical parameters as:
\begin{linenomath*}
\begin{equation}
\label{new_Cc}
\mathcal{C}_c =\alpha_c\mathcal{C}_T = \frac{8\pi}{15}.\frac{\rho{_c}\mathcal{R}^{5}_{CMB}}{\mathcal{MR}^2_T}.(1+\frac{2}{3}f_c)
\end{equation}
\end{linenomath*}
where $\mathcal{R}_{CMB}$, $\rho_c$, $\mathcal{M}$ and $\mathcal{R}_T$ are the CMB radius, core density, the lunar mass and lunar radius, respectively.
\subsection{Hydrostatic core-mantle boundary model} A three-layer model of the Moon consisting of a lunar crust, mantle and fluid core were considered. Using constraints of mass from a previous estimate \cite{Viswanathan2018} and iteratively obtained MoI, the non-spherical deviations of each layer interface were estimated by limiting the non-hydrostaticity to the lunar crust (see SI).

\section{Data analysis and regression}
\subsection{Data processing} The processing (or reduction) of LLR data requires a precise light-time computation with accurate modeling of geophysical and relativistic effects, well described in previous studies \cite{IERS2010,Viswanathan2018}. These refined reduction models enable a precise determination of the intrinsic distance information measured by the two-way time-of-flight of the laser pulses between one of the seven Earth stations and one of the five lunar retro-reflectors. A reduction model for LLR data \cite{Viswanathan2018} has been implemented within ``G\'eod\'esie par Int\'egrations Num\'eriques Simultan\'ees'' (GINS), an orbit determination and processing software of the ``Centre National d'\'Etudes Spatiales'' (CNES), validated through a step-wise comparative study.
\subsection{LLR-fit} 
\begin{figure}
\centering
\includegraphics[width=1.0\textwidth]{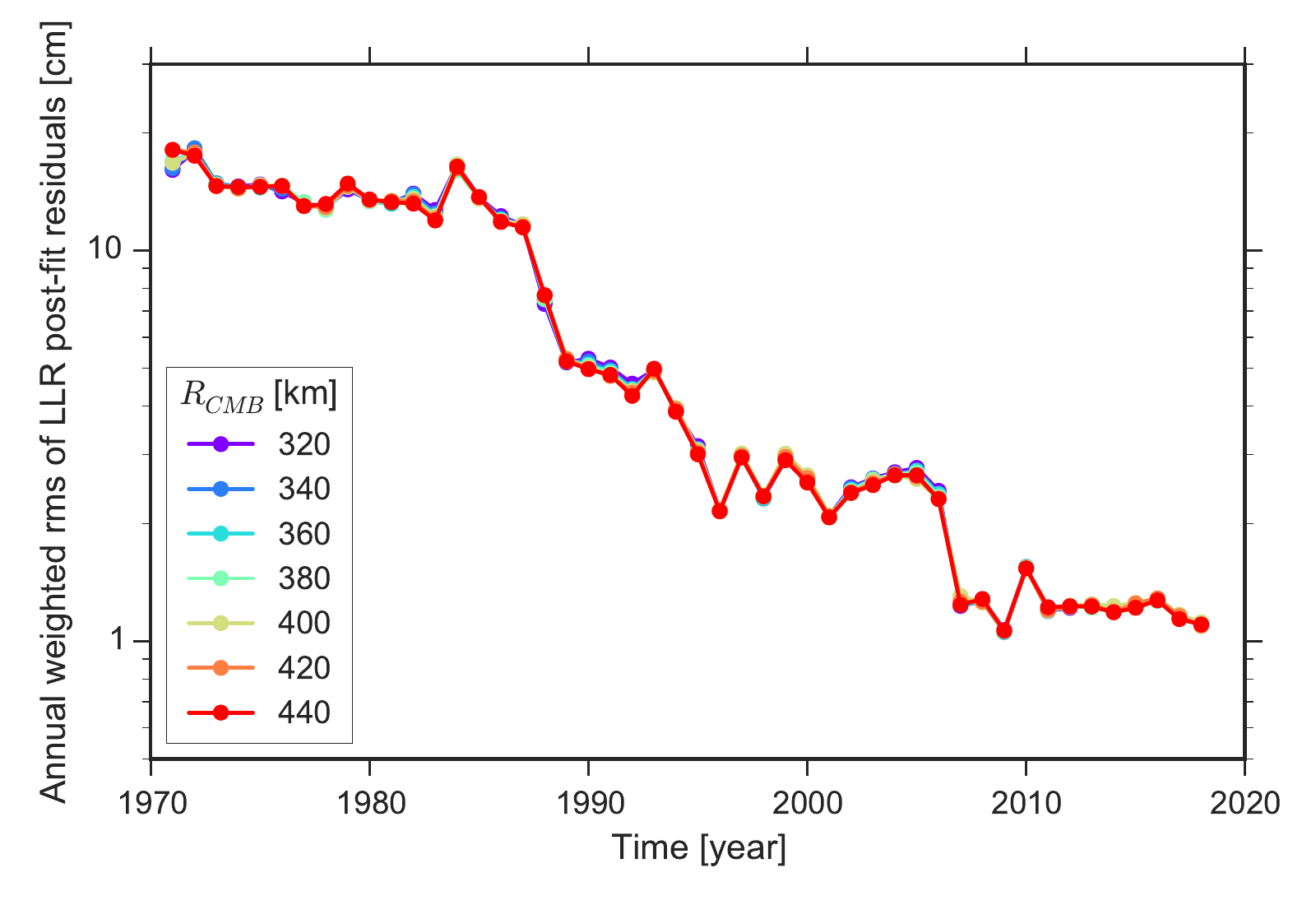}
\caption{Annual weighted root-mean-square (wrms) of LLR post-fit residuals obtained with the redefined dynamical model with the radius of the lunar core-mantle boundary ($\mathcal{R}_{CMB}$) varying between 320 to 440 at step-sizes of 20 km. The variations in wrms of LLR post-fit residuals between the solutions are well below the $\approx$ 5 mm observational accuracy of the LLR dataset. The downward trend indicates an improvement in the observational accuracy of LLR dataset by a factor of 20 over nearly 5 decades of observing.}
\label{img_residuals}
\end{figure}

The fit of the lunar part of the ephemeris to LLR data involves solving for parameters linked to the Earth-Moon orbit and rotation. A full list of adjusted and fixed parameters relevant to the fit are provided in Table \ref{table_param}. CMB radii between 320 to 440 km with a step-size of 20 km was chosen to be explored. For each CMB radius, the appropriate values of the density, shape and radius were chosen from the lunar interior model (see SI) and an iterative fit to the LLR solution was performed. The solution parameters were adjusted according to a chosen type (fit, fixed or constrained) mentioned in Table \ref{table_param}, and the annual weighted rms of the post-fit residuals are given in Fig.~\ref{img_residuals}.
\subsection{Uncertainty of $f_c$} The formal uncertainties obtained from the least-square fit were too small to be considered realistic. Hence, the impact of correlated parameters that are fixed or constrained (e.g., $\rho_c$, $d_{22,c}$, etc.), on the estimates of core oblateness were tested. This includes analyzing the impact of variations (at known uncertainties) of these parameters on the solutions, to quantify the relative error introduced on $f_c$ (see Table \ref{table_fc_sig}). A possible error from the non-linear behavior of the partial derivatives of $\boldsymbol{\omega_{c}}$ is quantified using a two-step process:
\begin{enumerate}
\item Iterative fit of initial conditions of $\boldsymbol{\omega_{c}}$ with partial derivatives obtained with their initial conditions set to zero.
\item Fit of initial conditions of $\boldsymbol{\omega_{c}}$ with partials obtained with new initial conditions (non-zero values) obtained from previous step.
\end{enumerate}
The above steps were performed for two CMB radii (340 and 420 km). The solutions obtained between these two sets of partial derivatives, impact the estimate of $f_c$ at a relative error of 0.4\%. The relative error introduced on $f_c$ from a fixed value of lunar mass is expected to be below 0.2\%, as the fractional uncertainty from the Newtonian gravitational constant ($\mathcal{G}$) is much larger than that from the lunar gravitational mass ($\mathcal{GM}= \mathcal{GM}_{EMB}/(1+EMRAT)$, where $\mathcal{GM}_{EMB}$ is the gravitational mass of the Earth-Moon barycenter and $EMRAT$ is the Earth-Moon mass ratio).
The error on the estimated value of $f_c$ (in Fig.~\ref{img_fc}) results from the cumulative variations of the estimated values of $f_c$ due to errors from fixed and constrained correlated parameters, tabulated in Table \ref{table_fc_sig} in the order of decreasing precedence.

\begin{sidewaystable}
\caption{Lunar parameters relevant to the fit of the dynamical and reduction model to LLR observations. Superscript ($\dagger$) represents fixed quantities.}
\label{table_param}
\begin{tabular}{lclcl}
Parameter 																					& Notation 																				& Note \\ \hline
Lunar Euler angles and their rates 												& $\phi,\theta,\psi,\dot{\phi},\dot{\theta},\dot{\psi}$ 			& initial condition (at J2000) \\
Core differential velocity 															& $\boldsymbol{\omega_{c}}$ 		 									& initial condition (at J2000) \\
Geocentric position and velocity of the Moon 							& $\boldsymbol{r_{EM}},\boldsymbol{\dot{r}_{EM}}$	& initial condition (at J2000) \\
Gravitational mass of E-M barycenter$^\dagger$ 						& $\mathcal{GM}_{EMB}$ 													& INPOP17a \cite{Viswanathan2018} \\
Earth-Moon mass ratio$^\dagger$ 												& $EMRAT$ 																			& INPOP17a \cite{Viswanathan2018} \\
Newtonian gravitational constant$^\dagger$ 								& $\mathcal{G}$ 																	& CODATA: 2014 \cite{Mohr2016}\\
Earth's orbital$^\dagger$(O) and rotational (R) time delay 		&$\tau_{O(0,1,2)}$,$\tau_{R(1,2)}$										& INPOP17a \cite{Viswanathan2018} \\
Lunar time delay for solid-body tide 											& $\tau_M$ 																			& - \\
Lunar gravity-field (up to deg-6)												& $C_{nm,T}, S_{nm,T}$ 													& within GRAIL uncertainties \cite{Konopliv2013} \\
																									& $C_{32,T}, S_{32,T}, C_{33,T}$ 									& adjusted below 1\% \cite{Williams2014b} \\ 
Lunar potential Love number 														& $k_2$ 																					& within GRAIL uncertainties \cite{Konopliv2013} \\
Lunar vertical displacement Love number 								& $h_2$  																				& - \\
Lunar horizontal displacement Love number$^\dagger$ 			& $l_2$  																					& model value of 0.0107 \cite{Williams2014b} \\
Polar MoI of the Moon 																& $\mathcal{C}_T/\mathcal{MR}^2_T$ 									& Eqn.~\ref{MoI_Moon} \\
Density of lunar core$^\dagger$ 												& $\rho_c$ 																				& 5000 to 7500 kg/m$^3$ (see SI) \\
Radius of core-mantle boundary$^\dagger$ 								& $\mathcal{R}_{CMB}$ 													& 320 to 440 km, 20 km steps \\
CMB polar shape coefficient 														& $d_{20,c}$ 																			& $f_c=-(3/2)d_{20,c}$ \cite{Meyer2011} (see SI) \\
CMB equatorial shape coefficient$^\dagger$ 							& $d_{22,c}$ 																			& hydrostatic value (see SI) \\
Euler angles for a non-principal axes CMB	$^\dagger$			& $\nu,\epsilon,\mu$ 														& Sensitivity test for non-zero off-diagonal core moments (see SI) \\
Coefficient of viscous friction at CMB 										& $K_v$ 														& -\\
Lunar retro-reflector (LRR) coordinates 									& $\boldsymbol{r}^{LRR}_{x,y,z}$ 									& 5 LRR: A15, A14, A11, L1, L2 \\
LLR station coordinates and velocities$^\dagger$						& $\boldsymbol{r}^{sta}_{x,y,z}$, $\boldsymbol{\dot{r}}^{sta}_{x,y,z}$				& INPOP17a \cite{Viswanathan2018} \\
LLR station biases																		& bias \#																					& INPOP17a \cite{Viswanathan2018}\\
\hline
\end{tabular}
\end{sidewaystable}

\begin{sidewaystable}
\caption{Impact of constrained model parameters on the estimated error on $f_c$ at $\mathcal{R}_{CMB}\approx381$~km. Relative error values provided are upper limits. The CMB equatorial shape was scaled up by a factor 5 of the reference model value following \citeA{LeBars2011}.}
\label{table_fc_sig}
\begin{tabular}{llccc}
\hline
Parameters               & References &Variations & Unit & Impact on $f_c$ \\
                                 &                 &                  &          & [rel. error \%] \\
\hline
Core density ($\rho_c$)  &\citeA{Garcia2011} & $\pm 1000 $ & kg/m$^3$  & $20$ \\
Crustal thickness ($T_{cr}$) &\citeA{Wieczorek2013}	& $34 - 43$  & km & $6$ \\
Potential love number ($k_2$) &\citeA{Konopliv2013} & $\pm 1.8 \times 10^{-4}$  & 1 & $3$ \\
Crustal density ($\rho_{cr}$) &\citeA{Wieczorek2013} & $\pm 18$ & kg/m$^3$ & $2$ \\
Tilt for a non-principal axes CMB ($\epsilon$) 	&\citeA{Wieczorek2019} & $6.4$  & deg & $1$ \\
CMB equatorial shape coefficient ($d_{22,c}$) & Reference model & $5 \times d_{22,c}$ & 1& $0.5$\\
Newtonian gravitational constant ($\mathcal{G}$) &\citeA{Mohr2016} & $\pm 3.1 \times 10^{-15}$ & m$^3$kg$^{-1}$s$^{-2}$ & $0.2$ \\
Mean lunar moment of inertia ($\overline{\mathcal{I}}$) &\citeA{Viswanathan2018} & $\pm 1 \times 10^{-5}$   & 1 & $0.1$ \\
\hline
\end{tabular}
\end{sidewaystable}

\begin{table}
\caption{Lunar interior model values obtained for $\mathcal{R}_{CMB}$ varying between 320 to 440 km in 20 km step-size.}
\label{table_values}
\begin{tabular}{ccccccccc}
\hline
 $\mathcal{R}_{CMB}$ &   $\rho_c$   &  $\alpha_c$    & $f_c$ & $f_{c,(hydrostatic)}$    &$\mathcal{C}_T/\mathcal{MR}^2_T$ &      $K_v/\mathcal{C}_T$ 	    &   $\tau_M$ & $h_2$ \\
  {[}km{]}			   & [kg/m$^3$] 	    & [$10^{-4}$]  & [$10^{-4}$]  & [$10^{-4}$] &    [1] & [$10^{-9}$ rad/day] & [day] 	  & [1] \\
\hline
    $320$ &  $7621.4 $& $ 4.92  $ & $3.76$ & $1.85$ & $0.39307$ & $7.06$ & $0.07534$ & $0.04334$ \\
    $340$ &  $6879.0 $& $ 6.01  $ & $3.13$ & $1.96$ & $0.39311$ & $7.32$ & $0.07519$ & $0.04333$ \\
    $360$ &  $6288.4 $& $ 7.31  $ & $2.60$ & $2.07$ & $0.39316$ & $7.55$ & $0.07503$ & $0.04328$ \\
    $380$ &  $5811.6 $& $ 8.85  $ & $2.17$ & $2.16$ & $0.39322$ & $7.72$ & $0.07506$ & $0.04332$ \\
    $400$ &  $ 5421.6$ &$ 10.66$ & $1.81$ & $2.24$ & $0.39329$ & $7.90$ & $0.07484$ & $0.04326$ \\
    $420$ &  $ 5098.1$ &$ 12.79$ & $1.51$ & $2.31$ & $0.39338$ & $8.02$ & $0.07471$ & $0.04336$ \\
    $440$ &  $ 4826.7$ &$ 15.28$ & $1.24$ & $2.37$ & $0.39348$ & $8.12$ & $0.07489$ & $0.04350$ \\
\hline
\end{tabular}
\end{table}

\acknowledgments
We acknowledge the ESEP (post-doctoral fellowship) and the PNGRAM for funding this research. The computations were performed on the servers of Geoazur - OCA and the LLR data processed using CNES-GRGS software (GINS). This work benefited from the previous contribution of H. Manche to the INPOP ephemeris development. We thank M. A. Wieczorek and two anonymous referees for their valuable comments that improved the manuscript. We acknowledge the continued efforts of personnel at Apache Point, Grasse and older stations (Haleakala, Matera and McDonald) for their respective contributions to the LLR dataset, archived by the International Laser Ranging Service \cite{Pearlman2002} at \url{https://ilrs.cddis.eosdis.nasa.gov/data_and_products/data_centers/index.html}


%
%

\bibliography{short_main}

%
%
%
%
%

\end{document}